\documentclass{aa}
\usepackage{graphicx}
\usepackage{txfonts}
\usepackage{footmisc}
\usepackage[]{natbib}
\bibpunct{(}{)}{;}{a}{}{,} % to follow the A&A style
\newcommand{\tcl}{RXCJ2359.3-6042}
\newcommand{\ftmsol}{$\times$10$^{14}$M$_{\odot}$}

\begin{document} 

\title{Witnessing a merging bullet being stripped \\
  in the galaxy cluster, \tcl}
\author{Gayoung Chon\inst{1}
  \and
  Hans B\"ohringer\inst{1}
}
\titlerunning{Discovery of a merging bullet in \tcl}
\authorrunning{Chon \& B\"ohringer} 
\institute{Max-Planck-Institut f\"ur extraterrestrische Physik,
  Giessenbachstrasse, Garching D-85748 Germany\\
  \email{gchon@mpe.mpg.de}
}
\date{Received xx 10, 2014; accepted 31 12, 2014}
\abstract
{
We report the discovery of the merging cluster, \tcl, from the REFLEX II cluster survey and present
our results from all three detectors combined in the imaging and spectral analysis of the XMM-Newton data. 
Also known as Abell 4067, this is a unique system, where a compact bullet penetrates an extended, low density 
cluster at redshift $z$=0.0992 clearly seen from our follow-up XMM-Newton observation. 
The bullet goes right through the central region of the cluster without being disrupted and we can clearly 
see how the bullet component is stripped of its layers outside the core. 
There is an indication of a shock heated region in the east of the cluster with a higher temperature.
The bulk temperature of the cluster is about 3.12($\pm$0.13)\,keV implying a lower mass system. 
Spearheading the bullet is a cool core centred by a massive early type galaxy. 
The temperatures and metallicities of a few regions in the cluster derived from the spectral analysis supports our conjecture 
based on the surface brightness image that a much colder compact component at 1.55($\pm$0.10)\,keV with large 
metallicity (0.75\,Z$_{\odot}$) penetrates the main cluster, where the core of the infalling component survived the merger 
and left stripped gas behind at the centre of the main cluster. 
We also give an estimate of the total mass within r$_{500}$, which is about 2\ftmsol\ from the deprojected spherical $\beta$ 
modelling of the cluster in good agreement with other mass estimates from the M--T$_{\rm X}$ and M-$\sigma_{\rm v}$ relations.
}
\maketitle
\section{Introduction}

Merging galaxy clusters have been found to be interesting laboratories for the study of a variety of astrophysical 
processes~\citep{feretti02,shock07}.
They are also the major process in which galaxy clusters grow and have therefore a large influence on the global 
properties of these systems. 
Observationally we see a wide range of different morphologies in cluster mergers, which is due to different merger 
geometries and the variety of morphologies of the merging clusters including different mass ratios.

Based on the complete flux-limited cluster survey REFLEX II in the southern sky~\citep{r2const} comprising more 
than 900 galaxy clusters, we constructed a volume-limited subsample (B\"ohringer et al., in prep.), in which we study 
108 of 157 clusters in XMM-Newton follow-up observations for a systematic study of galaxy cluster structure 
statistics~\citep{hbsub10,chon12}. 

In this survey we found a very interesting merging cluster, \tcl, also known as Abell 4067, at a redshift of 0.0992
with a special merger configuration that has never been observed in this form before. 
In this system we witness the encounter of two very dissimilar systems, a smaller compact system with a cool component, 
which penetrates a larger, more diffuse cluster with low central density in a bullet-like fashion. 
The fact that a compact bullet merges with an extended cluster without a dense core resembles the case of 
1E 0657-558, known as the Bullet Cluster~(\cite{bullet}).
Unlike this cluster, however, the relative velocities and the Mach number associated with \tcl\  are 
much lower, and so the interaction processes are quite different.
In particular, the interaction seems to be less violent and probably as a result we clearly see a trail of 
debris behind the bullet which is not observed in the Bullet Cluster. 
This can provide interesting information on the stripping process acting on the infalling component. 
The mass of the cluster and the associated ICM temperature are also lower than in the Bullet Cluster making 
temperature measurements easier. 
These circumstances make this object unique as a laboratory for the study of the stripping process as explained below.

Throughout the paper we adopt a flat $\Lambda$-cosmological model with $\Omega_{\rm m}$=0.3 and a Hubble constant 
of H$_0$=70~km\,s$^{-1}$Mpc$^{-1}$ for the calculations of all distant dependent quantities. 
We used N$_{\rm H}$ of 1.41$\times$10$^{20}$cm$^{-2}$ for the spectral analysis.
All quoted uncertainties refer to one sigma error.

\section{XMM-Newton observations}

\begin{figure}[ht]
  \centering
  \resizebox{\hsize}{!}{\includegraphics{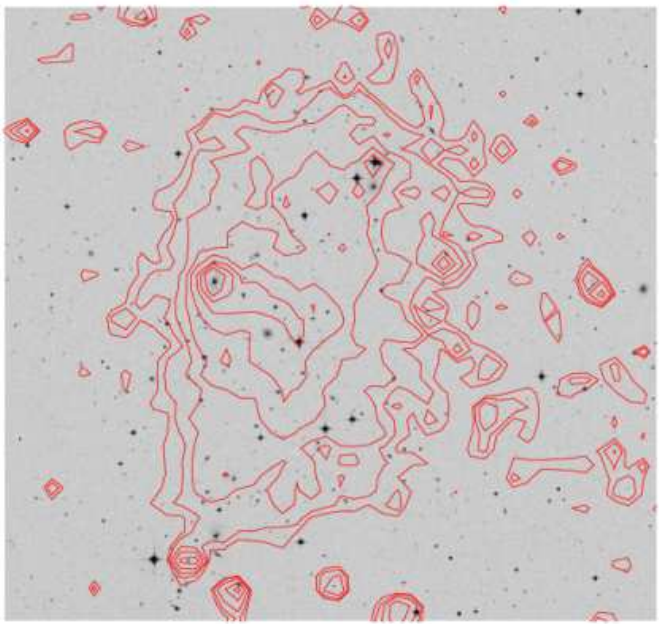}}
  \vspace{-0.3cm}
  \caption{
    \tcl\ shown in the DSS optical image overlaid with the XMM-Newton X-ray contours in the [0.5-2]\,keV band
    from the same data as in Fig.~\ref{fig:ov}. 
    The contours are obtained from a background subtracted image where the first contour traces 
    3$\times$10$^{-5}$\,counts~s$^{-1}$\,pix$^{-1}$. 
    The flux of the contours increases by steps of 1.5 subsequently. 
    The compact component with high X-ray emission is centred at the galaxy at 23$^h$59$^m$04.2$^s$ in R.A. 
    and -60$^d$36$^m$34$^s$ in Dec.
  }
  \label{fig:opt-xray}
\end{figure}

The digitised sky-survey (DSS) optical image is shown with the {\it XMM-Newton} X-ray contours in Fig.~\ref{fig:opt-xray}, 
which gives a first impression of the geometric configuration of \tcl. 
On top of the extended X-ray emission from the cluster we see a smaller, much stronger emission from a source, which 
coincides with an optical galaxy, near the centre of the image with an extended tail towards the south-west. 
This galaxy is at $z$=0.100~\citep{teague90}.

We used the XMM-Newton observation of \tcl\ with Obs.ID 0677180601 as a part of the REFLEX II cluster survey 
project to investigate further. 
After the flare cleaning both MOS have 11\,ks exposure while pn is left with 8\,ks. 
Standard data analysis was performed with SAS v13.0.1.
Our basic X-ray imaging data reduction follows closely the procedures outlined in~\cite{chon12sub}, and we briefly 
summarise them here. 
The XMM-Newton observation for all three detectors were flare-cleaned and out-of-time (oot) events were statistically 
subtracted from the pn data.  
Point sources were removed, and the modelled background was subtracted based on the blank sky observation provided
by~\cite{read03}. 
The cleaned images from all three detectors were combined with an exposure correction. 

For the spectral analysis we used Filter Wheel Closed (FWC) data to remove the contribution from the particle background. 
The FWC spectrum was scaled according to the ratio of the corner events between the observation and FWC data. 
We also removed the oot events of the FWC-pn data before scaling, which mostly affects the very low energy 
range\footnote{\footnotesize Newly added FWC oot pn data are found at 
\url{http://xmm2.esac.esa.int/external/xmm_sw_cal/background/filter_closed/pn/FF/pn_FF_2014_v1.shtml}}.  
The region outside $r_{500}$\footnote{\footnotesize The radius within which the mean cluster mass density is 500
  times the critical density of the Universe. To calculate r$_{500}$ for \tcl\ we used the recipes as described
  in~\cite{r2const}.} of the cluster within the field of view was taken as a background 
from which the contributions from the X-ray cosmic background together with the residual particle background were  
determined. 
We considered an unabsorbed thermal model for the Local Hot Bubble, unresolved point sources described by an 
absorbed power law, and a cool absorbed thermal halo. 
The unabsorbed thermal component is best fitted with 0.09\,keV while the absorbed thermal model with 0.29\,keV. 
In both cases the metallicity is set to unity and the redshift to zero. 
The power law describing unresolved point sources is best fitted with an index of 1.41. 
The residual particle background was modelled with a power law, 
where the best fit indices for MOS1, MOS2 and pn are 0.1, 0.41, and 0.46. 
The spectral regions were then fitted with a thermal component representing the cluster emission together with 
the background components described above using the APEC model in XSPEC. 
Given the limited photon statistics the values of all background components were fixed to the pre-determined values 
from the background region except for the normalisations, which are scaled appropriately according to the areas. 
There are known cross-calibration issues among the EPIC cameras~\citep{read14}; however, this is less of 
a problem for cooler clusters (see e.g.~\citet{xmmcal14}). 
Hence, in the final fitting the temperature, metallicity, and normalisation of the cluster were kept free, 
and all three detectors were linked apart from the normalisation.

\section{Interpretation of the data}

\subsection{Image analysis}

\begin{figure}[ht]
  \resizebox{\hsize}{!}{\includegraphics{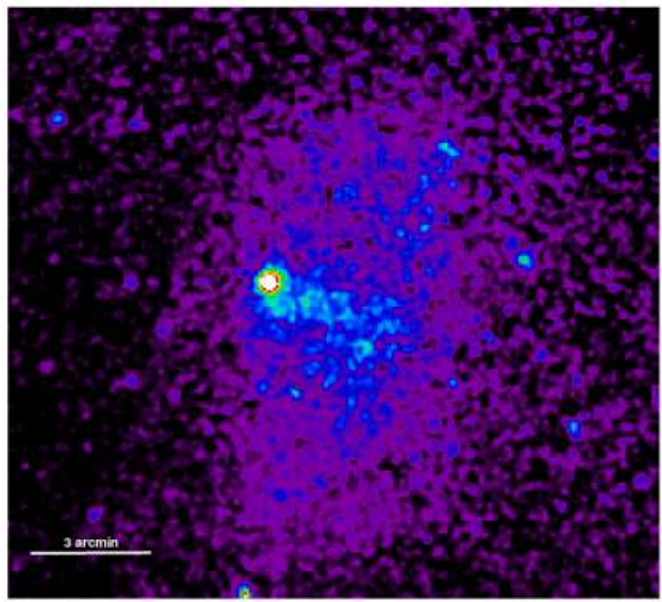}}
  \caption{
    XMM-Newton flux image of \tcl. All three detectors were combined after 
    the exposure correction in the [0.5-2.0]\,keV band. 
    A scale of three arcmin is shown with a bar.
  }
  \label{fig:ov}
\end{figure}

Figure~\ref{fig:ov} shows a combined flux image of three XMM-Newton detectors of the cluster in 
the [0.5-2]\,keV band, smoothed with 4$"$ kernel.

\begin{figure}
  \centering
  \resizebox{\hsize}{!}{\includegraphics{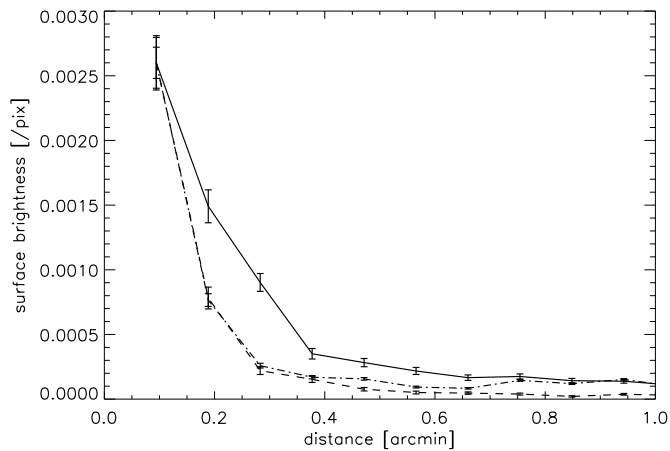}}
  \caption{
    Profiles of the cool and extended component penetrating the cluster (solid) 
    in the [0.5-2.0]\,keV band in comparison to two point sources in the same observation.
    One point source is located at a similar off-axis angle (dash-dotted) and the other point source 
    is at about twice the off-axis angle (dashed) as the cool component. 
  }
  \label{fig:extent}
\end{figure}

The main cluster shows a very shallow, extended X-ray surface brightness distribution. 
It appears as a dynamically young system which has not virialised, and has not formed a dense core. 
Alternatively, one can attribute the shallow surface brightness distribution to the disturbance caused 
by the merger itself, but in any case the main cluster could not have been a very regular cluster with 
a dense core before the merger, as we will argue below. 
Embedded in this low surface brightness cluster we see a small, bright compact component with a trail of patchy 
X-ray emission behind.
We extracted a radial surface brightness profile of this component, shown in Fig.~\ref{fig:extent} as a solid line. 
This profile is calculated only from the leading edge, which is almost identical to that of the entire region of 
the compact component. 
This confirms that the extended emission does not result from the debris trailing from the compact component. 
The centre of the profile is taken at the maximum of the X-ray emission, which coincides with the optical galaxy 
described in Sect. 2. 
Clearly it shows extended emission, which can be compared to the radial profiles of typical point sources in the 
same observation shown as dashed and dash-dotted lines. 
The most obvious interpretation of Fig.~\ref{fig:ov} is then that a compact cool core cluster has fallen into the diffuse, 
extended main system, and the trail of gas seen in the middle of the picture is a mixture of the gas stripped from 
the infalling bullet and main cluster. 
Inspection of the optical image in Fig.~\ref{fig:opt-xray} shows that the bright dominant galaxy is still at the centre 
of the infalling compact cluster. 
With the 18 galaxy redshifts available in the NED\footnote{\footnotesize The NASA/IPAC Extragalactic Database (NED) is 
operated by the Jet Propulsion Laboratory, California Institute of Technology, under contract with the National 
Aeronautics and Space Administration.} within the r$_{500}$ of \tcl\ we infer the approximate direction of the 
infall. 
For a cluster mass of 2\ftmsol , as will be estimated in Sect. 4, the  expected infall velocity is about 1310\,km\,s$^{-1}$. 
The line-of-sight relative velocity difference between the infalling bright core with respect to the cluster redshift 
is about 430\,km\,s$^{-1}$. 
This implies that the infall direction is close to the plane of the sky within about 20 degrees.
This configuration allows a relatively simple interpretation of the merging process limiting the ambiguity of 
projection effects. 

The aforementioned features reveal a particular case that the ``bullet'' has gone through the centre of the cluster, 
and still has preserved its core. 
In this case we expect the metallicity of this core region to be large, which indicates that it was undisturbed for 
long time~\citep{boehringer04}. 
From this we can further infer that main cluster did not have a dense core itself, otherwise the collision of the 
two cores would have disrupted the infalling core.
We can further deduce that if the main cluster did not have a cool core, the patchy emission that we observe as a trail of 
the ``bullet'' must be the debris of the outer layers of the infalling system.

Further inspection of the image shows a sharper edge of the cluster on the east side than on the western 
counterpart, and faint emission outside the east edge. 
Analogous to the bullet cluster, we expect the sharp edge (approximately bounded by the western sides of the green 
rectangular regions in Fig.~\ref{fig:region}) to be the contact discontinuity of the bullet material. 
In this case we would expect to see a shock further to the east of this edge,  
but because of the low surface brightness and poor photon statistics we cannot detect a surface brightness jump that 
would reveal the location of this shock.
Nevertheless, as another consequence of such a shock, we would expect the shock heated region to the east of the 
contact discontinuity to be quite hot.
An obvious difference to the Bullet Cluster is that we have a very shallow opening angle of the cone of the contact 
discontinuity which is expected to follow the shape of the possible Mach cone. 
This implies a very mild shock. 
The possible shock wave still needs to be located and needs to be better quantified together with the cold front
with better data in the future.

\subsection{Spectral analysis}

In this section we will study the temperature and metallicity distribution in the cluster, which will help to verify
our speculations about the history of the merger laid out in the previous section.

\begin{figure}
  \resizebox{\hsize}{!}{\includegraphics{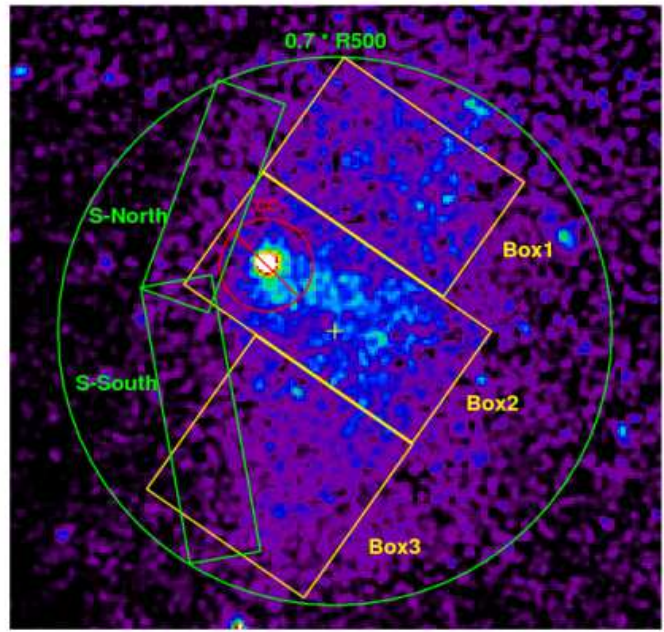}}
  \caption{
    Defined regions for the spectral analysis. 
    The central 0.7$\times$r$_{\rm 500}$ is marked by a green circle centred at the yellow cross 
    along with three boxes that divide the main cluster. 
    The centre of the cluster is at 23$^h$58$^m$52.3$^s$ in R.A. and -60$^d$37$^m$59$^s$ in Dec.
    The cool component (CC) is marked by a red circle centred at the highest X-ray emission. 
    The suspected shocked regions are enclosed by two green boxes.
  }
  \label{fig:region}
\end{figure}

Figure~\ref{fig:region} defines the regions from which we extracted the spectra. 
The cluster bulk properties are determined from the X-ray emission inside the large circle enclosing 0.7$\times$r$_{500}$ 
excluding the circular region centred at the infalling core, which was analysed separately. 
Given the extent of the cool component (CC) found in the surface brightness profile, the CC region was defined by a 
radius of one arcmin. 
The main part of the cluster is divided into three boxes; Boxes 1 and 3  are considered to be less affected 
by the infalling component than Box 2. 
We also excluded the CC from Box 2 when extracting the spectrum. 
The suspected shock-heated regions are marked by two green rectangles. 
The corresponding best fit temperatures and metal abundances for these regions are given in Table~\ref{tab:temp}. 

\begin{table}[htb]
  \centering
  \begin{tabular}{|c|c|c|}
    \hline
    region & T (keV) & Z$_{\odot}$ \\
    \hline
    CC & 1.55 $\pm$ 0.10 & 0.75 $\pm$ 0.19 \\
    cluster (-CC) & 3.12 $\pm$ 0.13 & 0.17 $\pm$ 0.04 \\
    Box1 & 3.54 $\pm$ 0.26 & 0.18 $\pm$ 0.09 \\
    Box2 (-CC) & 3.08 $\pm$ 0.12 & 0.27 $\pm$ 0.08 \\
    Box3 & 3.51 $\pm$ 0.22 & 0.18 $\pm$ 0.07 \\
    shock (N+S) & 4.5 $\pm$ 0.59 & 0.35 $\pm$ 0.22 \\
    \hline
  \end{tabular}
  \caption{{\footnotesize
      Best fit temperature and metallicity of the spectral areas defined 
      in Fig.~\ref{fig:region} with 1$\sigma$ errors. 
    }}
  \label{tab:temp}
\end{table}

The temperature measured within 0.7$\times$r$_{\rm 500}$ is 3.12\,keV with a metallicity of 0.17\,Z$_{\odot}$  without the CC. 
The bright, compact CC is found to be at the lowest temperature in the system with the largest metal abundance. 
This nicely confirms our conjecture about this region being an old cool core. 
The next coldest component is found in Box 2, which we assumed to be the debris of the infalling system mixed 
with the main cluster. 
Boxes 1 and 3 contain relatively hot gas in comparison. 
In particular, we find the region marked in green rectangles on the eastern front, which we assume to be shock-heated, 
to be the hottest region in the cluster. 
This is analogous to the shock heated region in the bullet cluster, but everything is scaled to lower temperature. 
Hence, the temperature and metallicity distributions support our conjecture about the merging process laid out in 
the previous section. 
Currently, we are limited by the photon statistics and can neither pin-point the exact location of the shocked region 
nor the contact discontinuity with the spectral analysis.

\section{Modelling the main cluster component}

To obtain more insight into the properties of the main cluster component, we use a spherical approximation 
to the shape of the cluster, excising a circular area around the CC and point sources. 
We fit a $\beta$-model with a constant background to the surface brightness distribution of the cluster 
whose centre, marked by a yellow cross in Fig.~\ref{fig:region}, is determined by the centre shift method 
described in~\cite{chon12sub}. 
The best fit value of $\beta$ is found to be 0.64$\pm0.062$, and that of the core radius to be 3.04$\pm0.32$ arcmin, 
both with 1$\sigma$ error. 
A variation of $\sim$10\% in the background corresponds to a change in the estimates of two parameters similar 
to the errors. 
With these values and the cluster bulk temperature found in the previous section we can estimate the 
total mass of the main cluster assuming a spherical symmetry and isothermality of the ICM, which allows an analytical 
deprojection of the observed surface brightness. 
We then get an estimate of the total mass of 
2.1($\pm$0.37)$\times$10$^{14}$M$_{\odot}$ for the cluster temperature of 3.54\,keV. 
We consider the temperatures in Boxes 1 and 3 of 3.54\,keV a better representation of the bulk temperature of
the undisturbed main cluster, compared to the temperature measurement in Box 2 or the total cluster region,
since the mixture of main cluster gas with colder gas stripped away from the bullet will bias the single temperature
fit low.
Nevertheless, to bracket the uncertainties in the temperature estimate, we also estimated the mass of the cluster 
for two temperature limits, 3 and 3.8\,keV. 
For this temperature range we find that the estimated mass lies between 1.8 and 2.2\ftmsol. 

It is also interesting to compare this mass estimate to other mass probes. 
With the estimated cluster temperature we can use the M--T$_{\rm X}$ scaling relation, for example by~\cite{arnaud05}, 
to obtain the cluster mass. 
Using the overdensity $\delta$ of 500 we get a mass estimate of 
2($\pm$0.25)$\times$10$^{14}$M$_{\odot}$ 
for the cluster temperature of 3.54\,keV. 
For 3.12\,keV we get a roughly 5.5\% lower mass, hence in both cases we have a good agreement with our hydrostatic mass estimate. 
We will consider another probe of mass in the next section with optical spectroscopic data. 

From the $\beta$-model fit to the cool core excised main cluster image, we determine a central electron density of about 
2.5$\times$10$^{-3}$\,cm$^{-3}$. 
This yields a central entropy value of 190$\pm$14~keV\,cm$^{-2}$, which is well in the regime of non-cool core clusters (see~\cite{pratt10}).
It is less extreme than that of Coma with its central entropy value of 535~keV\,cm$^{-2}$~\citep{coma13}, and that of 
the low surface brightness cluster A76 with a value of 400~keV\,cm$^{-2}$~\citep{ota13},  
but it shows that the properties of the main cluster is far from the stage where it could have developed a cool core, 
in agreement with our conclusion above.

\subsection{Dynamical mass estimate}

An estimate of the dynamical mass of a cluster can be inferred from measurements of spectroscopic redshifts of member galaxies.
In the presence of a large number of redshifts of a hundred or more a detailed analysis of the redshift distribution 
is appropriate as outlined by~\citet{beers90}, for example. 
For smaller data sets with at least ten redshifts a scaling relation of velocity dispersion and cluster mass 
as provided for example by Biviano et al. (2006) is considered a reasonable and robust approach. 
This calibration of the scaling relation is based on the study of 64 galaxy clusters in cosmological N-body simulations.
They consider that mass estimates are reliable for a sample that involves at least ten galaxy redshifts within the virial radius. 
There are 19 spectroscopic redshifts available in the NED database for \tcl\ within r$_{500}$, with which we calculated 
the velocity dispersion based on~\cite{harrison74} and~\cite{biviano06}. 
We applied a three-sigma clipping, which removes one galaxy, to obtain the final dispersion. 
We followed the recipes given in~\cite{biviano06} to calculate a mass estimate considering their Eq. 2
and Fig. 4, and obtain a mass estimate of 2.2\ftmsol\ for the one-dimensional velocity dispersion of  
605$\pm$87\,km\,s$^{-1}$ and $h$=0.7. 
The quoted error on the velocity disperson results from 1000 bootstrap realisations, which were subject to 
the same procedure as were the data,
which results in a 43\% error for the mass estimate including the uncertainty of the scaling relation
given in~\cite{biviano06}. 
Since the dynamical mass estimate is based on the dispersion of radial velocities, fair agreements between different 
mass estimates of this system implies that the dynamical disturbance cannot be large in the radial direction. 
It provides another signature that the merger is close to the plane of the sky, in agreement with the angular trajectory 
of the bullet component that we calculated earlier.

\section{Summary and outlook}

We presented the analysis of the XMM-Newton data of a merging system, \tcl, which shows a main cluster 
with a low X-ray surface brightness being penetrated by a cooler compact component whose core is
still intact. 
Our conjecture about the merging configuration is supported by the spectral temperatures and metal abundances 
found in different regions of the cluster. 
We find that the mass of the main cluster without the infalling component is about 2\ftmsol, which is 
based on three different mass estimates for a bulk temperature of 3.5\,keV. 
Within the current limit of photon statistics we can tentatively infer that the infalling component 
must have induced a mild shock given the wide opening angle. 
A deeper observation will offer a chance to quantify  the processes ongoing in this special object in greater detail. 
In particular, it may allow us to identify the exact location of the contact discontinuity, to obtain more quantitative details
about the shock, and to reconstruct some of the properties of the infalling system such as a limit to the total gas mass.

\begin{acknowledgements}
  We thank the XMM-Newton user support team, especially Ignacio de la Calle and Maria Santos-Lleo 
  for their support with the calibration issues. We thank the referee for the comments. 
  This work is based on observations obtained with XMM-Newton, an ESA science mission with instruments 
  and contributions directly funded by ESA Member States and the USA (NASA).
  This research has made use of the NASA/IPAC Extragalactic Database (NED) which is operated by 
  the Jet Propulsion Laboratory, California Institute of Technology, under contract with the National 
  Aeronautics and Space Administration. 
  We acknowledge support from the DfG Transregio Program TR33, the Munich 
  Excellence Cluster ``Structure and Evolution of the Universe'' and Deutsches Zentrum f\"ur Luft- 
  und Raumfahrt (DLR) with the program 50 OR 1403. 
\end{acknowledgements}

\footnotesize{
  \bibliographystyle{aa}
  \bibliography{paper}

\begin{thebibliography}{19}
\expandafter\ifx\csname natexlab\endcsname\relax\def\natexlab#1{#1}\fi

\bibitem[{{Arnaud} {et~al.}(2005){Arnaud}, {Pointecouteau}, \&
  {Pratt}}]{arnaud05}
{Arnaud}, M., {Pointecouteau}, E., \& {Pratt}, G.~W. 2005, \aap, 441, 893

\bibitem[{{Beers} {et~al.}(1990){Beers}, {Flynn}, \& {Gebhardt}}]{beers90}
{Beers}, T.~C., {Flynn}, K., \& {Gebhardt}, K. 1990, \aj, 100, 32

\bibitem[{{Biviano} {et~al.}(2006){Biviano}, {Murante}, {Borgani}, {Diaferio},
  {Dolag}, \& {Girardi}}]{biviano06}
{Biviano}, A., {Murante}, G., {Borgani}, S., {et~al.} 2006, \aap, 456, 23

\bibitem[{{B{\"o}hringer} {et~al.}(2013){B{\"o}hringer}, {Chon}, {Collins},
  {Guzzo}, {Nowak}, \& {Bobrovskyi}}]{r2const}
{B{\"o}hringer}, H., {Chon}, G., {Collins}, C.~A., {et~al.} 2013, \aap, 555,
  A30

\bibitem[{{B{\"o}hringer} {et~al.}(2004){B{\"o}hringer}, {Matsushita},
  {Churazov}, {Finoguenov}, \& {Ikebe}}]{boehringer04}
{B{\"o}hringer}, H., {Matsushita}, K., {Churazov}, E., {Finoguenov}, A., \&
  {Ikebe}, Y. 2004, \aap, 416, L21

\bibitem[{{B{\"o}hringer} {et~al.}(2010){B{\"o}hringer}, {Pratt}, {Arnaud},
  {Borgani}, {Croston}, {Ponman}, {Ameglio}, {Temple}, \& {Dolag}}]{hbsub10}
{B{\"o}hringer}, H., {Pratt}, G.~W., {Arnaud}, M., {et~al.} 2010, \aap, 514,
  A32

\bibitem[{{Chon} \& {B{\"o}hringer}(2012)}]{chon12}
{Chon}, G. \& {B{\"o}hringer}, H. 2012, \aap, 538, A35

\bibitem[{{Chon} {et~al.}(2012){Chon}, {B{\"o}hringer}, \& {Smith}}]{chon12sub}
{Chon}, G., {B{\"o}hringer}, H., \& {Smith}, G.~P. 2012, \aap, 548, A59

\bibitem[{{Feretti} {et~al.}(2002){Feretti}, {Gioia}, \&
  {Giovannini}}]{feretti02}
{Feretti}, L., {Gioia}, I.~M., \& {Giovannini}, G., eds. 2002, Astrophysics and
  Space Science Library, Vol. 272, {Merging Processes in Galaxy Clusters}

\bibitem[{{Fusco-Femiano} {et~al.}(2013){Fusco-Femiano}, {Lapi}, \&
  {Cavaliere}}]{coma13}
{Fusco-Femiano}, R., {Lapi}, A., \& {Cavaliere}, A. 2013, \apjl, 763, L3

\bibitem[{{Harrison}(1974)}]{harrison74}
{Harrison}, E.~R. 1974, \apjl, 191, L51

\bibitem[{{Markevitch} {et~al.}(2004){Markevitch}, {Gonzalez}, {Clowe},
  {Vikhlinin}, {Forman}, {Jones}, {Murray}, \& {Tucker}}]{bullet}
{Markevitch}, M., {Gonzalez}, A.~H., {Clowe}, D., {et~al.} 2004, \apj, 606, 819

\bibitem[{{Markevitch} \& {Vikhlinin}(2007)}]{shock07}
{Markevitch}, M. \& {Vikhlinin}, A. 2007, \physrep, 443, 1

\bibitem[{{Ota} {et~al.}(2013){Ota}, {Fujino}, {Ibaraki}, {B{\"o}hringer}, \&
  {Chon}}]{ota13}
{Ota}, N., {Fujino}, Y., {Ibaraki}, Y., {B{\"o}hringer}, H., \& {Chon}, G.
  2013, \aap, 556, A21

\bibitem[{{Pratt} {et~al.}(2010){Pratt}, {Arnaud}, {Piffaretti},
  {B{\"o}hringer}, {Ponman}, {Croston}, {Voit}, {Borgani}, \&
  {Bower}}]{pratt10}
{Pratt}, G.~W., {Arnaud}, M., {Piffaretti}, R., {et~al.} 2010, \aap, 511, A85

\bibitem[{{Read} {et~al.}(2014){Read}, {Guainazzi}, \& {Sembay}}]{read14}
{Read}, A.~M., {Guainazzi}, M., \& {Sembay}, S. 2014, \aap, 564, A75

\bibitem[{{Read} \& {Ponman}(2003)}]{read03}
{Read}, A.~M. \& {Ponman}, T.~J. 2003, \aap, 409, 395

\bibitem[{{Schellenberger} {et~al.}(2014){Schellenberger}, {Reiprich},
  {Lovisari}, {Nevalainen}, \& {David}}]{xmmcal14}
{Schellenberger}, G., {Reiprich}, T.~H., {Lovisari}, L., {Nevalainen}, J., \&
  {David}, L. 2014, ArXiv e-prints

\bibitem[{{Teague} {et~al.}(1990){Teague}, {Carter}, \& {Gray}}]{teague90}
{Teague}, P.~F., {Carter}, D., \& {Gray}, P.~M. 1990, \apjs, 72, 715

\end{thebibliography}
}

\end{document}